\documentclass[runningheads]{llncs}
\usepackage{graphicx}
\usepackage{cite}
\usepackage{rotating}
\usepackage{enumerate}
\usepackage{wrapfig}
\usepackage{paralist}
\usepackage{booktabs}
\usepackage{url}

\usepackage{color}
\usepackage{ifthen}
\newboolean{showcomments}
\setboolean{showcomments}{true} 
\ifthenelse{\boolean{showcomments}}
    {
        \newcommand{\per}[1]{\textcolor{blue}{{\it [Per says: #1]}}}
        \newcommand{\johan}[1]{\textcolor{cyan}{{\it [Johan says: #1]}}}
    }
    {
        \newcommand{\per}[1]{}
        \newcommand{\johan}[1]{}
    }

\begin{document}
%

\title{How to Enable Collaboration in Open Government Data Ecosystems: A Public Platform Provider's Perspective}
%
\titlerunning{How to Enable Collaboration in Open Government Data Ecosystems}
%
\author{Johan Lin{\aa}ker\inst{1} \and Per Runeson\inst{1}}
%
\authorrunning{J. Lin{\aa}ker and P. Runeson}
%
\institute{Lund University, Ole Römers väg 3, Lund, Sweden
\email{\{johan.linaker,per.runeson\}@cs.lth.se}\\}
\maketitle              
\begin{abstract}
\textit{Background:} 
Open Government Data (OGD) is an important driver for open innovation, improved efficiency, and transparency for public entities. 
However, extant research highlights a need for improved feedback loops, collaboration, and a more demand-driven publication of OGD.


\textit{Objective:} 
Our objective is to explore how public entities in the role of platform providers can address this issue by enabling collaboration within their OGD ecosystems, both in terms of the OGD published on the underpinning platform, as well as any related Open Source Software (OSS) and standards.
\textit{Method:} 
We conducted an exploratory multiple-case study of four OGD ecosystems with diverse characteristics. Data was collected through semi-structured interviews, and in one of the cases through a prolonged engagement. The data was then coded using a set of \textit{apriori} codes. 
\textit{Results:} 
The study descriptively presents each case based on the coding, along with synthesis in the form of a conceptual model that highlights different attributes of OGD ecosystems. For example, we observe how collaboration can be enabled through different types of ownership of the platform provider, how the ecosystem's scope can vary, what roles the platform provider may undertake, how to enable open collaboration, and how to collaborate in terms of data sharing, OSS development, and standards. For each aspect, we provide recommendations based on the explored cases that, together with the model, may help public entities in designing and orchestrating new or existing OGD ecosystems.
\textit{Conclusions:} 
We conclude that enabling and facilitating collaboration in an OGD ecosystem is a complex exercise, yet believe that it offers new ways for public entities in how they can leverage the power of open innovation to address their goals and directives.



\end{abstract}

%
%


\section{Introduction}
\label{sec:introduction}

Rich and high-quality data have over time become a critical asset for software organizations as a driver for innovation and input to solutions including artificial intelligence~\cite{munappy2019data, gao2020generating}. One way of increasing access and availability of such data is to share it as Open Data~\cite{attard2015systematic} and collaborate on its collection and maintenance as commonly done with Open Source Software (OSS)~\cite{munir2015open}. Such sharing of data is less common within the software industry~\cite{RunesonNIER2019}, but more so among public entities~\cite{safarov2017utilization}. In the latter case, we refer to the openly shared data as Open Government Data (OGD)~\cite{attard2015systematic}. More specifically, we consider OGD as \textit{``\ldots government-related data that is made open to the public''}~\cite{attard2015systematic} and thereby a subset of Open Data.


To enable the potential innovation output from OGD~\cite{Lakomaa2013, janssen2012benefits}, public entities along with cross-sector organizations and citizens may form OGD ecosystems. These function as value networks through which the OGD is enriched and used as input to new or improved products and services~\cite{lindman2015business, zuiderwijk2014innovation}. The interaction and collaboration in these ecosystems are however often quite limited~\cite{m2017open, immonen2014requirements, dawes2016planning}, even though it is highlighted as a need~\cite{sieber2015civic, rudmark2019harnessing} and implicit characteristics of a data ecosystem~\cite{oliveira2019investigations}.

Considering OSS ecosystems (commonly referred to as communities)~\cite{franco2017open}, open collaboration on the underpinning OSS projects is common practice. By working together in an open, transparent, and egalitarian manner~\cite{feller2002understanding}, new functionality is continuously asserted, discussed, and implemented~\cite{ernst2012case}, thereby accelerating the innovation and development beyond what any single actor within the ecosystem could perform alone~\cite{munir2015open}.

OGD ecosystems may, potentially, have an opportunity to improve the innovation output and enable further value creation by adopting such practices~\cite{linaaker2020public}. Also, considering open data ecosystems, such as OpenStreetMap\footnote{\url{https://www.openstreetmap.org/}} and Wikidata\footnote{\url{https://www.wikidata.org/}}, there may also be a potential for catalyzing both the sharing and adoption of OGD by extending collaboration to include open development of boundary resources~\cite{dal2014role} such as related standards and software as OSS~\cite{zuiderwijk2014innovation, immonen2014requirements, rudmark2020open}. As we consider current definitions of data ecosystems~\cite{oliveira2019investigations} (including related concepts~\cite{susha2020towards}) lacking in terms of these characteristics, we define a data ecosystem, adapted from three sources~\cite{zuiderwijk2014innovation, jansen2020focus, oliveira2019investigations}, as:

\textit{ 
    a networked \emph{community of actors} (organizations and individuals), which base their relations to each other on a \emph{common interest}~\cite{zuiderwijk2014innovation}, 
    supported by an underpinning \emph{technological platform}~\cite{jansen2020focus}
     that enables actors to process data (for example, find, archive, publish, consume, or reuse) as well as to foster innovation, create value, or support new businesses~\cite{oliveira2019investigations}. 
     Actors \emph{collaborate on the data and boundary resources} (for example, software and standards), through the exchange of information, resources, and artifacts~\cite{jansen2020focus}.}


For OGD ecosystems specifically, the provider of the technological platform is constituted by a public entity (or part of one). Further, we consider any data as OGD, which is published by the public entity constituting the platform provider, whether it is produced by the public entity itself, or collected from or shared by an actor within the ecosystem. These actors may, for example, include public entities, firms, non-profit organizations, research institutions, and citizens. 




Existing research has mainly focused on the collaborative practices used in OSS ecosystems~\cite{alves2017software, jansen2020focus}, and has in terms of OGD ecosystems~\cite{attard2015systematic} (and similar concepts~\cite{susha2020towards}) been limited, both regarding collaboration on OGD or any related software or standard, even though identified as a need~\cite{sieber2015civic, oliveira2019investigations, rudmark2019harnessing}. The research goal of this study is therefore to \textit{explore how public entities in the role of platform providers can enable collaboration within their OGD ecosystems, both in terms of the OGD published on the underpinning platform, as well as any related boundary resources in the form of OSS and standards}. 

We take the perspective of the public entity constituting the platform provider as these manage the platform and the ecosystem around it. The platform provider lays out the development direction of the platform, decides what is shared, and establishes the governance structure~\cite{baars2012framework}, i.e., rules and processes for how the ecosystem can influence and collaborate on the development and any shared resources (for example, data, software, or standards). We find this as an interesting perspective as trust towards the platform provider is pivotal to enable collaboration and growth of an ecosystem~\cite{susha2020towards, CoyleValue2020}. 

To address the goal of this study, we conducted a multiple-case study~\cite{runeson2012casestudy} of four OGD ecosystems in which OSS, standards, and related collaborative practices are adopted, aiming to foster collaboration and increase the adoption of OGD. This is an extension of our previous work~\cite{linaaker2020collaboration} where we reported on findings related to two of the four cases presented in this paper. The four ecosystems are governed by public entities in the role of platform providers and focus on OGD related to the labor market in Sweden, public transport in Sweden, and the region of Helsinki in Finland, as well as the City of Chicago, Illinois, USA. Based on our findings we synthesize and present a conceptual model that describes how OGD ecosystems may vary in terms of different aspects, for example including, platform ownership and scope, as well as sharing and collaboration practices. For each aspect, we provide recommendations based on the explored cases that, along with the conceptual model, may help public entities in designing and orchestrating new or existing OGD ecosystems with the ambition to enable collaboration with and within the ecosystems.

The rest of this paper is structured as follows. In section~\ref{sec:background} we provide a background on software and OGD ecosystems and their governance structures. In section~\ref{sec:researchDesign} we present our research design, followed by the descriptive results from our case studies in section~\ref{sec:results}. In section~\ref{sec:analysis}, we present and discuss our synthesis from the case studies in the form of a conceptual model along with a number of recommendations. Finally, we summarize and conclude the paper in section~\ref{sec:conclusions}.

\section{Background}
\label{sec:background}
The frame of our analysis and discussion of the studied cases is set by literature on software and data ecosystems, and particularly their organizational and governance structures, as outlined below.

\subsection{Software and Data Ecosystems}

Stemming from the literature on business and digital ecosystems~\cite{jansen2009sense, iansiti2004keystone, oliveira2019investigations}, software and data ecosystems share a common background, yet they differentiate in many ways.


Software Ecosystems~\cite{alves2017software, jansen2020focus} offer a lens for analyzing how networked communities of organizations collaborate around their common interest in a central software technology~\cite{mhamdia2013performance}. The central software technology, commonly referred to as a platform~\cite{jansen2009sense}, 
can be viewed as a \textit{``foundation technology or set of components used beyond a single firm and that brings multiple parties together for a common purpose or to solve a common problem''}~\cite{gawer2002platform}. The platform provider enables the ecosystem to use and extend the functionalities of the underpinning platform by providing boundary resources, i.e., \textit{``the software tools and regulations that serve as the interface for the arm's‐length relationship between the platform owner and the application developer.''}~\cite{ghazawneh2013balancing} Such resources can be of both technical (for example, API:s, developer tools, and libraries) and social character (for example, documentation, forums, and hackathons)~\cite{dal2014role}.


A Data Ecosystem is defined by Oliveira et al.~\cite{oliveira2019investigations} as a \textit{``socio-technical complex network in which actors interact and collaborate with each other to find, archive, publish, consume, or reuse data as well as to foster innovation, create value, and support new businesses}. When comparing the definitions of software and data ecosystems~\cite{jansen2009sense, oliveira2019investigations}, similar characteristics can be recognized, for example, in terms of actors, their relationships, as well as use, sharing, and collaboration of common resources~\cite{oliveira2019investigations}. One difference however is that this definition of a data ecosystem lacks the mentioning of an explicit underpinning technological platform and any associated boundary resources. This observation further adds to other definitions of data ecosystems surveyed by Oliveira et al.~\cite{oliveira2019investigations}. 

Instead of software platforms, the focus in data ecosystems is rather on the flow of data between the actors in the ecosystem in the form of a value-chain~\cite{lindman2015business}. A general distinction between the roles in such value-chains can be made between data providers and data users~\cite{zuiderwijk2014innovation}. The roles can be further refined into data providers, service providers, data brokers, application developers, application users, and infrastructure and tool providers~\cite{immonen2014requirements, kitsios2017business}. The data provider is usually constituted by a public-sector organization~\cite{oliveira2019investigations} as data sharing from private actors is not as widespread~\cite{RunesonNIER2019}. Services or functions needed include an infrastructure to share the data (preferably from multiple providers), documentation, tools for application developers, help in finding use-cases, as well as the possibility to discuss, provide feedback and make requests~\cite{immonen2014requirements, zuiderwijk2014innovation, dawes2016planning}. Several researchers observe the need for improved feedback-loops, collaboration, and a more demand-driven publication of OGD~\cite{rudmark2019harnessing, zuiderwijk2014innovation, dawes2016planning, m2017open}.



In this study, we introduce an adapted definition of a data ecosystem (see section~\ref{sec:introduction}) where we explicitly highlight the presence of an underpinning technological platform as well as associated boundary resources. Considering the definition of a platform by Gawer and Cusumano~\cite{gawer2002platform}, we consider the platform as the technological mean that in part or full enables the ecosystem’s actors to process data (for example, find, archive, publish, consume, or reuse) as well as to foster innovation, create value, or support new businesses, inspired by Oliveira et al.~\cite{oliveira2019investigations}. With the definition, we also look specifically at Open Government Data (OGD) ecosystems, i.e., data ecosystems with a public actor (or part of one) providing a platform for sharing OGD~\cite{attard2015systematic}.

\subsection{Ecosystem Organizational and Governance Structures}
\label{sec:background:ecosystemRolesAndGovernanceStructrues}

OGD ecosystems are typically centered around \textit{``government organizations as central actors, taking the initiative within networked systems organized to achieve specific goals related to innovation and good government''}~\cite{harrison2012creating}. 
Oliveira et al.~\cite{oliveira2019investigations} propose a classification of ecosystems' organizational structures, two of which highlight the position of the platform provider, either as an \emph{orchestrator} and main data provider of the ecosystem, or more limited as an \emph{intermediary} between data producers and users of the ecosystem. 


Considering software ecosystems specifically with the definition by Jansen et al.~\cite{jansen2009sense}, these structures blend and overlap. Dal Bianco et al.~\cite{dal2014role} mainly differentiate between two organizational structures; \textit{keystone-centric} ecosystems where the platform is controlled by a central independent organization, and \textit{consortium-based} ecosystems where the central organization is created and co-owned by the ecosystem's members. 

Looking further, three types of roles are commonly referred to~\cite{jansen2020focus, iansiti2004keystone}. The first role is that of the \textit{platform provider} who is the owner and supplier of the platform and thereby also usually the orchestrator of the ecosystem. As an orchestrator, the platform provider also decides on the governance model for the ecosystem, meaning how it maintains control and decides on the direction, but also on the governance structure, meaning ``the distribution of rights and responsibilities among the [ecosystem's members], and the rules and protocols that need to be followed to make decisions regarding the [ecosystem]''~\cite{alves2017software}. 

\textit{Keystone} and \textit{niche players} are two other roles within an ecosystem. A keystone is an actor who nurtures a symbiotic relationship with the ecosystem and its other actors, looking to actively improve its health~\cite{jansen2014measuring}. Usually, they have a close connection with the platform provider, who also may be referred to as a keystone if it has similar symbiotic intents. Niche players are actors focused more on a specific niche of the market, or use-case, and is primarily a user of the resources provided by the ecosystem~\cite{iansiti2004keystone}.

For OSS ecosystems, the platform provider can be the owner of the OSS project, usually either a software vendor or the ecosystem of actors directly or via a proxy organization (for example, a foundation)~\cite{riehle2012single}. Governance, however, does not have to be aligned with the ownership. In more autocratic ecosystems, it can be centered around a vendor or an individual, while in more democratic ecosystems it is distributed~\cite{de2013evolution}. In the latter case, control of the OSS project is usually maintained by a central group of actors who have gained a level of influence by proving merit, building trust, and social capital through contributions to the OSS project.

\section{Research Design}
\label{sec:researchDesign}
This study presents an exploratory multiple-case study~\cite{runeson2012casestudy} that we conducted to investigate four instances of OGD ecosystems, see Table~\ref{tbl:cases}. Case 1 (JobTech Dev) was selected due to the first author’s prolonged engagement in its platform provider. Cases 2--3 (Trafiklab and HSL DevCom)  were selected due to their characteristics as what we consider as mature examples of OGD ecosystems (based on our provided definition), but also due to their Nordic context where the government has a central role and high level of trust. Case 4 (The City of Chicago) was selected purposely to get an ecosystem with a general domain (municipality) compared to the other examples of labour market and public transport, and also to provide an example beyond the Nordic context.

The actors in the four ecosystems involves organizations and individuals from both public and private sectors, civic society, academia, and private citizens.
The unit of analysis is the ecosystems' \textit{collaboration} around OGD published on the ecosystems' underpinning platforms along with any related OSS and standards.

\begin{table}[]
\caption{Characteristics of the four OGD ecosystems studied.}
\label{tbl:cases}
\begin{tabular}{p{0.8cm} p{2.5cm} p{2.5cm} p{1.5cm} p{1.5cm} p{2.5cm}}
\toprule
\textbf{Case} & \textbf{Ecosystem} & \textbf{Platform provider} & \textbf{Scope} & \textbf{Founded} & \textbf{Owners} \\ \midrule
1 & JobTech Dev & Swedish\newline Public\newline Employment Service (SPES) & Swedish \newline labor market & 2018 & Swedish national government \\
2 & Trafiklab & Samtrafiken & Swedish \newline
public transport & 2011 & Regional public transport authorities and privately owned transport operators \\
3 & HSL DevCom & Helsinki\newline Regional\newline Transport\newline authority (HRT) & Helsinki\newline regional\newline public\newline transport & 2009 & Municipalities in the region \\
4 & City of\newline Chicago & City of\newline Chicago & City of\newline Chicago & 2011 & City of\newline Chicago \\ \bottomrule
\end{tabular}
\end{table}

To be able to analyze the collaboration, we used a model from our earlier work~\cite{linaaker2020collaboration} to visualize roles and structure of the studied ecosystems' governance. The model is commonly used for describing the governance in OSS ecosystems~\cite{nakakoji2002evolution} and was recently applied to a government-initiated OSS ecosystem~\cite{robles2019setting}.


The research effort was initiated with Case 1, where the first author of this study is embedded as an action researcher, as a part of a long-term research project. The researcher was hence able to generate in-depth knowledge through prolonged engagement along with access to extensive documentation. The documentation, field notes and three semi-structured interviews are used to triangulate findings (for questionnaire, see supplementary information\footnote{\url{https://github.com/johanlinaker/supplementary-material-jedem-2020}}). To ensure construct validity~\cite{runeson2012casestudy}, we based the questionnaire on earlier work about assessing the governance structure of software ecosystems~\cite{alves2017software, jansen2020focus}. The inteviewees were the platform's product owner, community manager, and policy strategist.

Data was gathered from Case 1 before any intervention was introduced from the action research. To avoid researcher bias, peer-debriefing between the first and second authors was performed continuously~\cite{runeson2012casestudy}.

\begin{table*}[t!] 
\centering
\caption{Interviewees from the four cases JobTech Dev, Trafiklab, HSL DevCom and City of Chicago.}
\label{tbl:interviewees}
\begin{tabular}{p{0.7cm} p{3.5cm} p{2.5cm} p{2.5cm} p{2cm}}
\toprule
\textbf{ID} & 
\textbf{Role} & 
\textbf{Ecosystem} & 
\textbf{Employer} & 
\textbf{Years with Ecosystem} \\ \midrule

I1 & 
Product owner & 
JobTech Dev &
SPES & 
2017--\\\midrule

I2 & 
Community Manager & 
JobTech Dev &
SPES & 
2018--\\\midrule

I3 & 
Policy Strategist & 
JobTech Dev &
SPES & 
2018--\\\midrule

I4 & 
Product Manager & 
Trafiklab &
Samtrafiken & 
2019--\\\midrule

I5 & 
Data specialist & 
HSL DevCom &
HSL & 
2010--\\\midrule

I6 & 
Unit Manager & 
City of Chicago &
City of Chicago & 
2013--2018 \\\bottomrule
\end{tabular}
\end{table*}

For Cases 2--4, data were also gathered through semi-structured interviews (see Table~\ref{tbl:interviewees}), using the same questionnaire.
All interviews were audio-recorded with additional notes taken. A threat regarding the reliability concerns that only the first author conducted the interviews~\cite{runeson2012casestudy}. To mitigate the threat, member-checking was performed in both cases, where synthesized findings were presented to all interviewees who were asked for correctness, misinterpretations, and redundancy. 

A series of \textit{apriori} codes were defined based on our definition of an OGD ecosystem as presented in Section~\ref{sec:introduction}.
\begin{itemize}
    \item[\textbf{C1}] Scope and purpose of ecosystem
    \item[\textbf{C2}] OSS projects shared as part of platform
    \item[\textbf{C3}] Open Data shared as part of platform
    \item[\textbf{C4}] Stakeholder and ecosystem actors
    \item[\textbf{C5}] Ecosystem governance
    \item[\textbf{C6}] Collaborative aspects related to OSS
    \item[\textbf{C7}] Collaborative aspects related to Open Data
    \item[\textbf{C8}] Collaborative aspects related to standards
\end{itemize}

Synthesized findings are presented per case in Section~\ref{sec:results} where codes have been abstracted into the following four categories:

\begin{itemize}
    \item General Background (\textbf{C1})
    \item Platform Content (\textbf{C2--3})
    \item Ecosystem Governance Structure (\textbf{C4--C5})
    \item Orchestration and Collaboration (\textbf{C6--C8})
\end{itemize}

Quotes from the interviewees (see table~\ref{tbl:interviewees}) are used to provide further context and nuance to the findings. The cases were then cross-analyzed based on the apriori codes which resulted in a conceptual model of OGD ecosystems. The model consists of eight aspects, each with a number of attributes, which emerged from the analysis. The model is visualized is Fig.~\ref{fig:collaborationFramework} and presented in detail in Section~\ref{sec:analysis}.

\section{Results}
\label{sec:results}

This section summarizes the results from the four studied cases, JobTech Dev, Trafiklab, HSL DevCom, and the City of Chicago.



\begin{figure}[t]
\begin{center}
\includegraphics[width=\columnwidth]{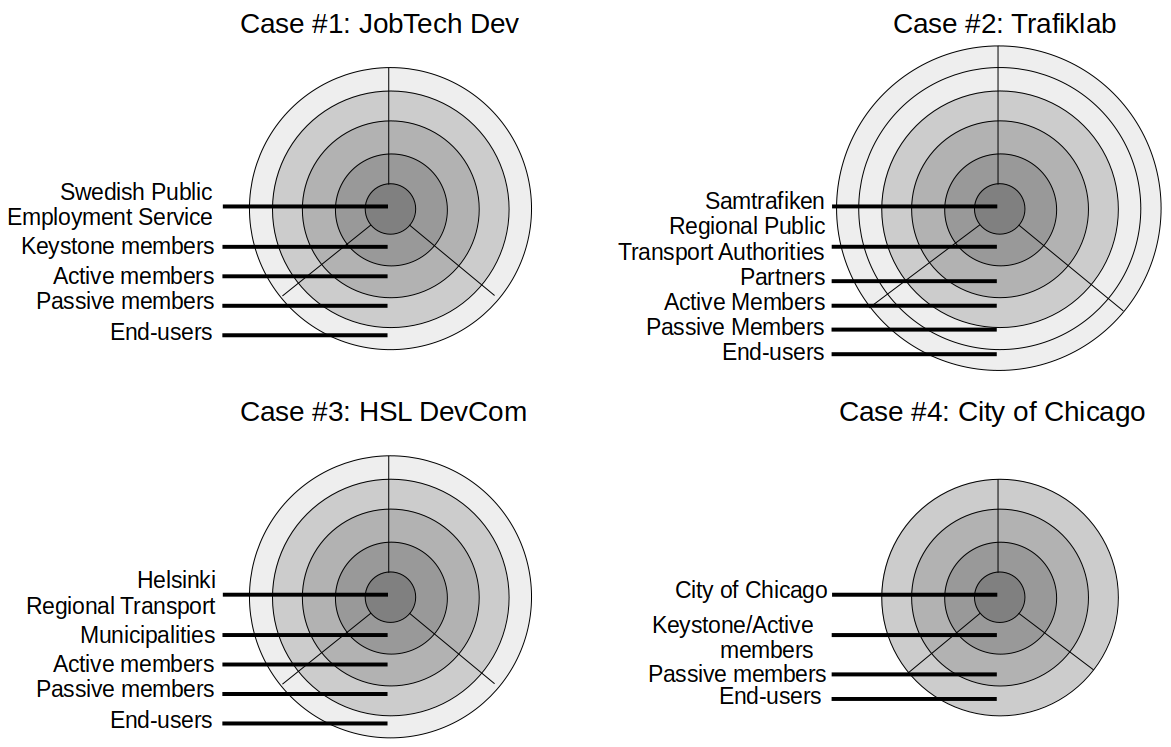}
\vspace{-0.5cm}
\caption{Overview of the governance structure of the four cases in this study, JobTech Dev (upper left), Trafiklab (upper right), HSL DevCom (lower left), and City of Chicago (lower right), based on the governance model presented in earlier work~\cite{linaaker2020collaboration} (in turn adopted from Nakakoji et al.~\cite{nakakoji2002evolution}). The core is occupied by the platform provider who controls development and data shared on platform. Ecosystem actors are divided among a number of layers specific for each ecosystem. Actors important to the platform provider and ecosystem, and with high influence on the platform are close to the core. Influence and activity decrease for each additional layer. The closer to the core, the more influence they have on the platform}
\vspace{-0.8cm}
\label{fig:Cases}
\end{center}
\end{figure}

\subsection{Case 1: JobTech Dev}

\subsubsection{General Background}
JobTech Dev is an ecosystem, initiated in 2018, bringing actors together which operate within or in relation to the Swedish labor market, to collaborate on a common platform of OGD, connected APIs, and complementary OSS projects. The ecosystem and its platform are developed and orchestrated by the Swedish Public Employment Service (SPES), a Swedish national government agency responsible for enabling the match-making between job-seekers and employers on the labor market. JobTech Dev was created as a part of this mission to enable the actors in the ecosystem to accelerate their innovation process, improve their services, and thereby improving the digital match-making on the labor market.

\subsubsection{Platform Content}
The platform consists of three parts: \textit{Jobs, Taxonomy}, and \textit{Career}.

Jobs is a service intended to collect job advertisements on the Swedish labor market through a collaboration with several ad providers and make the ads available through a common API.

Taxonomy is a collection of skills and job titles and relationships between them. The data set is developed and maintained within SPES. By opening up the data through APIs, the actors in the ecosystem are enabled to ``speak the same language'' enabling, for example, improved reporting and statistics and match-making between job advertisements and job-seekers' resumes. The data is released under the CC0 1.0 license\footnote{\url{https://creativecommons.org/publicdomain/zero/1.0/}}.

Career is (unlike Taxonomy and Jobs) not OGD. It is a service where job-seekers can store their resumes on a central location in an encrypted format. The job-seekers can grant and withdraw permission to organizations, for example, recruitment firms, social networks, and insurance firms to access their information. 
The service is based on the MyData principles\footnote{\url{https://mydata.org/guiding-principles/}} and enables job-seekers to only have to maintain one copy of their resume and to distribute and manage their data with kept control over their integrity and privacy.

The source code for the APIs is available as OSS as are the algorithms. To lower adoption barriers to the data, example applications are also developed and released as OSS. Members consuming the data, specifically startups, have acknowledged the value provided by these examples as it helps them understand use-cases and accelerate their development. 

\subsubsection{Ecosystem Governance Structure}
In terms of operations, the members of the ecosystem can generally be categorized within one of the areas: 1) recruiting and staffing firms, 2)~education and guidance providers, 3)~national, regional or local governments, 4)~workers' unions, 5)~employers' associations, 6)~job advertisers, and 7)~job seekers. Depending on the category, a member's interest in the platform may be limited to certain parts of the platform.

To visualize the governance structure, we use a governance model presented in earlier work~\cite{linaaker2020collaboration} (in turn adopted from Nakakoji et al.~\cite{nakakoji2002evolution}). SPES is positioned in the center as the platform provider orchestrating and governing the ecosystem (see Fig.~\ref{fig:Cases}, upper left). SPES ultimately decides on requirements and road-map for the platform, including what data to make available, when, and how. They perform the necessary development and maintenance and provide the infrastructure needed to access and use the data. 

In the layer closest to SPES are the members whose opinions may be considered as extra important for SPES in terms of developing the platform and growing the ecosystem. These members may, for example, have a large user base, or valuable competencies and resources, and thereby contribute to the health of the ecosystem. In the second layer are the general members and in layer three the end-users. Each layer is viewed to potentially consist of members from all types of operations. 


\subsubsection{Orchestration and Collaboration}
Due to the limited internal resources, SPES cannot maintain formal and direct relationships with all ecosystem members.
Teams carrying out development inside SPES therefore primarily work and communicate through close relationships with the key members to optimize the impact. However, SPES is striving to adopt an open development model and maintain an open dialogue where the whole ecosystem (including all layers in Fig~\ref{fig:Cases}) can influence the direction of the platform.

Anyone can, for example, request and discuss a new feature, an API, or data-set through an asynchronous open communication platform or by attending occasional meetups arranged by SPES. It is also possible to contribute to the development, although external contributions so far have been limited to bug reports and feature requests. The intention however is to encourage and enable members to contribute both new projects and to exist. Non-code contributions are however more common, for example, in the form of validating the quality and performance of the OSS. One example constitutes algorithms used to remove duplicate job ads as well assign statistical identifiers to the job ads that are collected and then made available through an API. External actors, including the Swedish statistics agency, have thus been able to test and validate the quality of the algorithms and whether they can trust the results the algorithms produce.

In terms of data, there are examples -- although limited -- of members producing and contributing directly to the platform. One example is a set of soft skills and their relationship to different job titles, which was contributed to the Taxonomy part of the platform. Processes are not yet established for how these types of contributions should be managed. In this case, a formal contract was established between the two parties. 

Close collaborations and direct dialogues with key members have been important to establish the ecosystem and gain general acceptance. For example, the collection of all job-ads on the market and making these freely available, made incumbents offering recruitment and staffing services initially question the intent from SPES as well as the suggested benefits. SPES views the commoditization of job-advertisement data, as with the ecosystem at large, as a way to push the actors forward who are working with digital match-making and guidance services, nurturing innovation and lowering barriers to entry for new actors.

After a more than two-year process, even the more conservative incumbents started to accept the ecosystem and see potential benefits with it. A formal collaboration was initiated between SPES and the ten largest job advertisement providers where they agreed to allow job-advertisements to be collected. Once collected, the ads are converted to an industry-specific open standard, and then enriched with metadata such as date of publishing and deadline for applications. A compromise was reached to only provide a ``stub'' of the advertisements, meaning that only that job title, metadata, and a link to the original advertisement would be provided through the platform's API.

SPES works actively to adopt or develop open standards when needed to promote and improve interoperability between the actors in JobTech Dev as well as towards the outside. One example is the adoption of the HROpen standard\footnote{\url{https://www.hropenstandards.org/}} that was adopted for the way individuals' resumé data is distributed. To also make their MyData-based implementation underpinning the Career-initiative, they have initiated an collaboration for defining interfaces and API:s within the MyData community\footnote{\url{https://mydata-infrastructure-project.github.io/}}.

\subsection{Case 2: Trafiklab}
\subsubsection{General Background}
Trafiklab is an ecosystem, initiated in 2011, that brings actors within the Swedish public transport sector together to collaborate on a platform with open traffic data, connected APIs, and complementary OSS projects. The ecosystem's vision is to facilitate the creation of new services that makes it easier and more attractive to travel with public transport. The ecosystem and its platform are developed and orchestrated by Samtrafiken, a corporate entity co-owned by all the regional public transport authorities and most of the commercial transport operators in Sweden. Operators also have the option of being partners to Samtrafiken without being co-owners.

\subsubsection{Platform Content}
The platform consists of data-sets and APIs, either maintained by Samtrafiken or independently by members of the ecosystem. All data hosted on the Trafiklab-platform is released with a custom license based on the principles of the CC0 1.0 license.

Four APIs provide static and real-time data on public transport, related to, for example, time-tables and interruptions. This data is currently made available in two types of standard formats, maintained by Samtrafiken and gathered from the regional public transport authorities and private operators following a government directive. Two further APIs provide time-table data for a trip-planner, an externally procured product that \textit{``\ldots is offered for free as-a-Service''} (I4) to the ecosystem.

Certain APIs are maintained by other organizations, both public and private, and made available on the Trafiklab platform. Data includes time-table and service data from Stockholm Public Transport and traffic information from the Swedish Transport Administration. The platform also links to related APIs that are maintained and hosted by other organizations. These include data from regional public transport authorities, local counties, and private entities.

In regards to OSS, they currently have a number of software development kits and example applications available. They intend to develop a new OSS trip planner and share their APIs as OSS along with the different parts of the platform.

\subsubsection{Ecosystem Governance Structure}
In terms of operations, the members of the ecosystem can generally be categorized within one of the areas: 1)~regional public transport authorities, 2)~private and publicly owned train operators, 3)~national, regional, and local governments, 4)~private bus operators, and 5)~private product and service providers. Plans include integration with related actors, such as taxi operators and rental-service providers of, for example, cars and bikes.

In terms of the governance structure, Samtrafiken is positioned in the center as the platform provider orchestrating and governing the ecosystem (see Fig.~\ref{fig:Cases}, upper right). Outside of Samtrafiken in the first layer are regional public transport authorities. As these are formal owners of the platform, they have a strong influence on the direction of Trafiklab. In the second layer are the formal partners to Samtrafiken which may include actors with different types of operation. The third layer primarily consists of private product and service providers, while end-users are positioned in the fourth layer.


\subsubsection{Orchestration and Collaboration}
Close relationships are maintained to regional public transport authorities and partners as these are the primary data producers but also consumers. The ecosystem at large can report bugs, ask for help, and request and discuss new features, APIs, or data-sets through an asynchronous, open communication platform. Physical meetings can also serve a similar purpose since Samtrafiken frequently hosts hackathons and meetups related to Trafiklab. 

As with SPES, Samtrafiken is transitioning to a more open and collaborative way of engaging with its ecosystem, a need identified in earlier research~\cite{rudmark2019harnessing}. What they have released thus far has mostly \textit{``\ldots been sporadic -- we need to have this for that meetup''} (I4). I4 adds however, \textit{``After this change we will have a more clear strategy regarding open source, where everything we do will be available as open source''}. One part in this are the \textit{``\ldots plans to build an open source journey planner based on the same data as the proprietary version''}~(I4). Another part in its adoption of more open and collaborative practices, is Samtrafiken move towards opening up its roadmaps, while also discussing a more formal approach where users can request and vote on what data sets should be prioritized. External contributions have thus far been limited to bug reports and feature requests.

Regarding the data, all of the provided data sets originate from data producers within the ecosystem. Depending on the case, Samtrafiken may transform the data to certain standard formats, develop and maintain the necessary APIs, and provide the necessary infrastructure for data consumers. 
A challenge with growing the ecosystem and gaining new data producers is related to standard formats of the data. For smaller actors, it is an expensive process to transform the data, and for Samtrafiken a recognized risk is that data may be destroyed when transformed between standards. Samtrafiken is, therefore \textit{``\ldots developing an input-portal to enable actors that do not have the resources or can afford it, to share their data on Trafiklab and to automate the transformation process''} (I4). The portal is specifically intended for actors in areas related to public transport, such as taxi operators and rental-service providers. The input-portal is a result of a long-term investigation conducted by Samtrafiken and its partners into the future potential and needs for public transport-related OGD. The investigation also rendered in a plan to introduce 12 new data sets by 2021. 

Other than helping data producers to transform their data into different standards (for example, the GTFS format, an industry standard for public transportation schedules and associated geographical information\footnote{\url{https://en.wikipedia.org/wiki/General_Transit_Feed_Specification}}), Samtrafiken and its partners within the Trafiklab ecosystem also collaborate on the development of new standards when needed. A standard for tickets and payment transactions was developed \textit{``\ldots with the goal of enabling sales of tickets across regional borders which was difficult to solve in existing proprietary solutions''} (I4).

\subsection{Case 3: HSL Developer Community}
\subsubsection{General Background}
HSL Developer Community (HSL DevCom) as an ecosystem was created in 2009 as the first data sets were published. However, \textit{``\ldots the ecosystem was far from established, for example, we did nothing related to open source software at this time''} (I5). The ecosystem brings together actors with an interest in OGD and OSS related to the Helsinki region's public transport (compared to the national focus in the case of Trafiklab). The ecosystem and its platform are developed and orchestrated by the Helsinki Regional Transport (HSL as per the Finnish translation) authority, a local joint authority co-owned by nine municipalities in the Helsinki region of Finland. HSL is, among other things, responsible for the planning and organizing the region's public transport system.

\subsubsection{Platform Content}
The platform consists of several data-sets and connected APIs as well as a number of related OSS projects central to the ecosystem actors, both in terms of internal operations and as customer-facing applications. All data is made available under the CC-BY 4.0 license\footnote{\url{https://creativecommons.org/licenses/by/4.0/}}.

The main part of the platform is related to the \textit{Journey Planner}\footnote{\url{https://digitransit.fi/en/}}, an OSS application for travelers planning their journeys within the region's public transport system. The Journey Planner was HSL's first OSS project piloted in 2013 where the core consists of the Open Trip Planner OSS project. \textit{``The user interface we have built ourselves, we use Open Trip Planner for the route planning and any outward-facing API:s''} (I5). The Journey Planner is developed in collaboration with the Finnish Transport Agency and Waltti, a public transport travel card collaboration in Finland. \textit{``HSL leads the development while the other two partners mainly contribute financially''} (I5). The platform further provides four underlying open APIs which provide \textit{i)} itinerary- and timetable data, \textit{ii)} geocoding data with addresses and coordinates, \textit{iii)} route suggestions between two coordinates, and \textit{vi)} the underlying map data. The data for \textit{i--iii} is produced by HSL while the map data for \textit{vi} is generated from OpenStreetMap (OSM), after a number of operations has been performed such as \textit{``filtering of the data, creation of grid patterns, and visualizations using [HSL's] own map style''} (I5).

A second main OSS application, also using the same underlying APIs, is the \textit{Map Generator}\footnote{\url{https://github.com/HSLdevcom/hsl-map-publisher}}, a web-based tool for generating, for example, route maps, traffic bulletins, and stop posters. The development of the Map Generator is mainly performed by HSL.

Other APIs that are provided via the platform covers data related to:
\begin{itemize}
    \item Bluetooth beacons installed on buses, bus stops, and terminals for trains and subways. One use case is to maintain an overview of the number of passengers waiting for or currently undergoing a ride. Another use case is to provide \textit{``\ldots better information regarding the current ride they are taking, and provide traffic information and potential alternative traveling options''} (I5). The data is produced and published by HSL.
    \item Trips made with city bikes in Helsinki. The data includes information related to the trips start and endpoints and is mainly used for analysis. The data is produced by City Bike Finland and contributed to HSL once a month who publish it on the platform.
    \item Car park information of commuters. The data is produced by municipalities and operators via a browser-based user interface on HSL's website. The data is then published on via the API on the platform by HSL.
\end{itemize}

\subsubsection{Ecosystem Governance Structure}
Similar to Trafiklab, members of the ecosystem can in terms of operations generally be categorized as: 1) national, regional, and local transport authorities 2) and governments, but also 3) private bus and 4) train operators, and 5) private product and service providers. As the use of the Journey Planner stretches beyond Finland, members of the ecosystem also cover Norwegian, Swedish, German, and Italian markets (to various extents).

In terms of the governance structure, HSL is positioned in the center as the platform provider (see Fig.~\ref{fig:Cases}, lower left). Outside in the first layer are the nine municipalities that co-own HSL and thereby have the formal ownership of the platform and a strong influence on the direction of its development where \textit{``\ldots the larger municipalities has a bit more influence''} (I5). In the second layer are the formal partners to HSL in relation to the Journey Planner, i.e., the Finnish Transport Agency and Waltti, which help to fund the development and maintenance of the Journey Planner and its underlying APIs. In the third layer are the active members of the ecosystem, which primarily include the national and international users of the Journey Planner, as well as large consumers of the data APIs such as Google who incorporates it in their search and map products. In the fourth layer are the more passive private product and service providers, while end-users are positioned in the fifth layer.

\subsubsection{Orchestration and Collaboration}
HSL has since the launch of the Journey Planner had an open-by-default perception, that most of what they develop should be available as OSS. \textit{``We had this proprietary solution we needed to replace, and we had thoughts on that we wanted to try doing it open source and use OpenStreetMap, then we found the Open Trip Planner project which does a lot of the things we need''} (I5). Hence, the authority has a large set of OSS projects available on their GitHub account. Among these, it is only the Journey Planner that benefits from external contributions. Due to deviations in interest and agendas, there is a branch of the Journey Planner that is being developed and maintained by the Norwegian Transport Authority. Both sides are however working towards integrating the two versions of the Journey Planner.

I5 reflects that the HSL has generally been focused on the development and hoped that the external contributions would come automatically with time, which, according to the interviewee only holds for the Journey Planner OSS project. However, in an upcoming large development project related to their master data registry, they have an \textit{``\ldots ambition to establish a collaboration with other public transport operators or organizations from start''} (I5).

In terms of the OSS projects, there are issue trackers for each specific project on GitHub. Bug reports or feature requests are however limited, with the Journey Planner as an exception. However, there is no general open issue tracker for the platform and its APIs that is actively used. Instead, the e-mail is the most common option along with HSL's social media accounts at Facebook and Twitter, which can be considered as the main communication channels used towards the ecosystem at large. They do use Slack, a synchronous communication platform which, however, is closed and exclusive for the main partners and international users of the Journey Planner OSS project. Physical meetings are frequently used to communicate with the ecosystem, present plans, and collect requirements, typically by speaking at conferences or partnering with external hackathons. 

In terms of data, there is a mix of data produced by HSL (for example, itinerary- and timetable data, and Bluetooth beacon data), by actors within the ecosystem (for example, geocode data from national authorities and car park data from the municipalities), and by ecosystems external to HSL DevCom (for example, map data from OpenStreetMap). Some data are automatically collected such as geocode data from municipalities or map data from OpenStreetMap, whilst other data may be shared by the ecosystem actor through technical infrastructure (car park data) or directly via mail (city bike trips data). 

To share maintenance and reduce the amount of internal edits needed, HSL has a full-time editor and some part-time resources available as needed that contribute to OpenStreetMap. Edits may, for example, include the move or addition of a bus stop, or the correction of street names. There have been contributions of larger data sets to OSM as well. These are however limited due to that HSL considers the current contribution process expensive in terms of resources, and due to conflicting licenses. \textit{``We have to provide the national road database with bus stop data licensed under CC-BY-4.0 and can therefore not base all of our data on OSM, because that would ''infect'' the data we export with OSM's [ODbL] license''}.

As with Samtrafiken, HSL promotes the use of the GTFS standard \textit{``\ldots after a request from Google, however, it was rather inevitable, as it has become a de facto format, even though there are other standards used as well''} (I5). No examples were given regarding development of new standards or specifications within the ecosystem.

\subsection{Case 4: City of Chicago OGD ecosystem}
\subsubsection{General Background}
The City of Chicago's OGD ecosystem was initiated in 2011 through the launch of the city's open data program and data portal. The ecosystem at large covers the whole city, including its citizens, NGOs, and companies. The main goal is to make the city's internal data useful to and consumable by the public. The first driver of the initiative was to increase transparency, which with time evolved to a higher focus on enabling social and economic development. 
Another important motive is to engage the public to improve their community. Citizens with a tech background see it as a way to \textit{``pay back to [the] community and be able to volunteer some of their time by using their skills''} (I6).

\subsubsection{Platform Content}
The platform consists of about 1400 different data sets available through API:s. The data springs out of the administrative processes that are carried out by the City of Chicago. There is hence a wide variety in the type of data spanning from every permit issued by the city, every crime reported, to every trip taken via mobility and ride-sharing operators. The most popular data set however was the \textit{``the list of salaries and all the employees who worked for the city''} (I6). The data is released under a custom licence\footnote{\url{https://www.chicago.gov/city/en/narr/foia/data_disclaimer.html}}.

In terms of OSS, the city has about 80 published OSS projects on their GitHub account, where a smaller number is considered more active and used than others. A common theme is that there is a connection to the data published on the city's data portal. Two of the more popular projects are made up of algorithms, that together with data from the portal, support decisions on where to send food inspectors in the city, and predicting E. coli levels in Lake Michigan, respectively. Another project regards \textit{OpenGrid}\footnote{\url{https://github.com/Chicago/opengrid}}, a web-based map that allows the user to explore multiple data sets simultaneously. 

\subsubsection{Ecosystem Governance Structure}
The members of the ecosystem surrounding the OGD and OSS provided via the city's platform are cross-sectoral, involving research institutions, companies, Non-Governmental Organizations (NGOs), journalists, and citizens. As there is a wide variety in the data sets provided there is no specific industry or area to classify these members in as with the other cases presented in this paper, such as actors connection to the Swedish labor market as in the case of JobTech Dev.

Considering the governance structure, the city is in the center position of the governance structure as the platform provider (see Fig.~\ref{fig:Cases}, lower right). At the layers closest, they don't prioritize certain groups before others (as with SPES for example) but listen to those actors that are actively communicating and collaborating with them. A representation can be found among all sectors listed above. Journalists, for example, is highlighted as one category who often interacts and requests the release of data sets related to their investigations. NGOs are another important type of actor as they often interface and reach larger groups of citizens either in general or around certain topics. Research institutions are also of importance as they are often part of developing OSS applications and data science projects using the published data. Certain companies, including startups and incumbents, may be more vocal than others depending on the their dependence of the data. Some citizens may also be considered as keystones by frequently filing public record requests.

As with the other cases, in the layers that follow are the more passive users of the data, followed by the end-users to the many applications that are created by the actors in the middle layers.

\subsubsection{Orchestration and Collaboration}
As with HSL, the City of Chicago has initiated a number of OSS projects that are available on the city's GitHub account. Of these, only parts are developed and maintained actively and are getting external contributions. In this open collaboration, research institutions make up important partners as they in turn are incentivized to collaborate with the government and also share the goals of the city in terms of \textit{``creating broad public value''} (I6) through the development of OSS. External grants have been a common tool to fund this collaboration and thereby the development of OSS projects such as OpenGrid. 

As highlighted in the OpenGrid project, also companies constitute important partners. Their contributions are in part development procured by the city, but also come as an effect as companies can offer and sell support for the OSS project to other customers (which was actively encouraged by the City of Chicago). Another successful project in terms of external contributions is the algorithm for measuring E. coli levels in Lake Michigan. The project, which was led by the city with a staff of eight people, received \textit{``over a thousand hours of in-kind contributions''} (I6). 

For OSS projects that are actively maintained, the city tries to publish both roadmaps and documentation, and use the collaborative infrastructure offered by GitHub. I6 describes how they with time have \textit{``figured out what sticks and then where you actually have to pay your attention to''} and also \textit{``take lessons from others and see how other [OSS] projects organize themselves}. For prioritized projects such as OpenGrid, the city organizes public meetings, physically and virtually, where the city can solicit feedback and present the roadmap so that \textit{``people can contribute to it and in a way that makes sense''} (I6). In certain cases, working committees would be set up to facilitate collaboration around specific topics. A common theme among the OSS projects is that they in some way make use of the data and enable analysis and exploration of it. The city views OSS as a mechanism for interacting with the citizens around the open data program and enable them to make use of the data, and also contribute back.

The data shared to large extents originate from within the city. However, it is not the city's intention \textit{\ldots to accumulate all data related to the City of Chicago} (I6). Instead, they try \textit{``\ldots to be pragmatic, add things of value [and] be able to meet people's needs''}  (I6). Hence, there are in general no specific domains that are considered in terms of sharing data. Sometimes there are exceptions when a specific initiative is highlighted, \textit{like clean energy and sustainability where some specific data sets would be released} (I6). Not all data originates from within the city, however. Some data sets are gathered from the national government, such as census data, population data, and socioeconomic data, and then broken down into the city's community areas. Mobility data generated from services such as taxi, ride-sharing, and bike rental is also collected and published due to legal requirements. 

The city has attempted to collaborate on certain data sets with the public by publishing the data sets on GitHub, similar to what is done with the source code in OSS projects. These data sets mainly connected to geographic entities such as \textit{``bike paths, streets, bike racks, [as] these are things very much in the public where [the citizens] actually might be able to see something and know it before the government knows about it''} (I6). One example of external contributions regarded footprints of the city's buildings. This data set was in turn contributed by citizens to the OpenStreetMap. Although successful examples, the city did \textit{``not experience GitHub as a good platform for sharing geospatial data and contributing data''} (I6) and also as the interest from the public was limited why they did not pursue this model of collaboration any further beyond the initial initiatives.

The city uses open standards when possible to enable interoperability between similar data sets from other cities, for example, through the use of GTFS related to public transport (similar to Samtrafiken and HSL) and the Mobility Data Specification related to mobility services. They also engage and co-develop new standards for how to publish data when needed. One example referred to a collaboration between a number of cities on how to commonly display the location and availability of flu shots.

\section{Analysis and Discussion}
\label{sec:analysis}
\begin{sidewaysfigure}
\caption{A conceptual model for OGD ecosystems consisting of eight main aspects and connected attributes as explained in section~\ref{sec:analysis}. Each attribute has the corresponding case number within brackets where it was identified.}
\includegraphics[width=\textwidth]{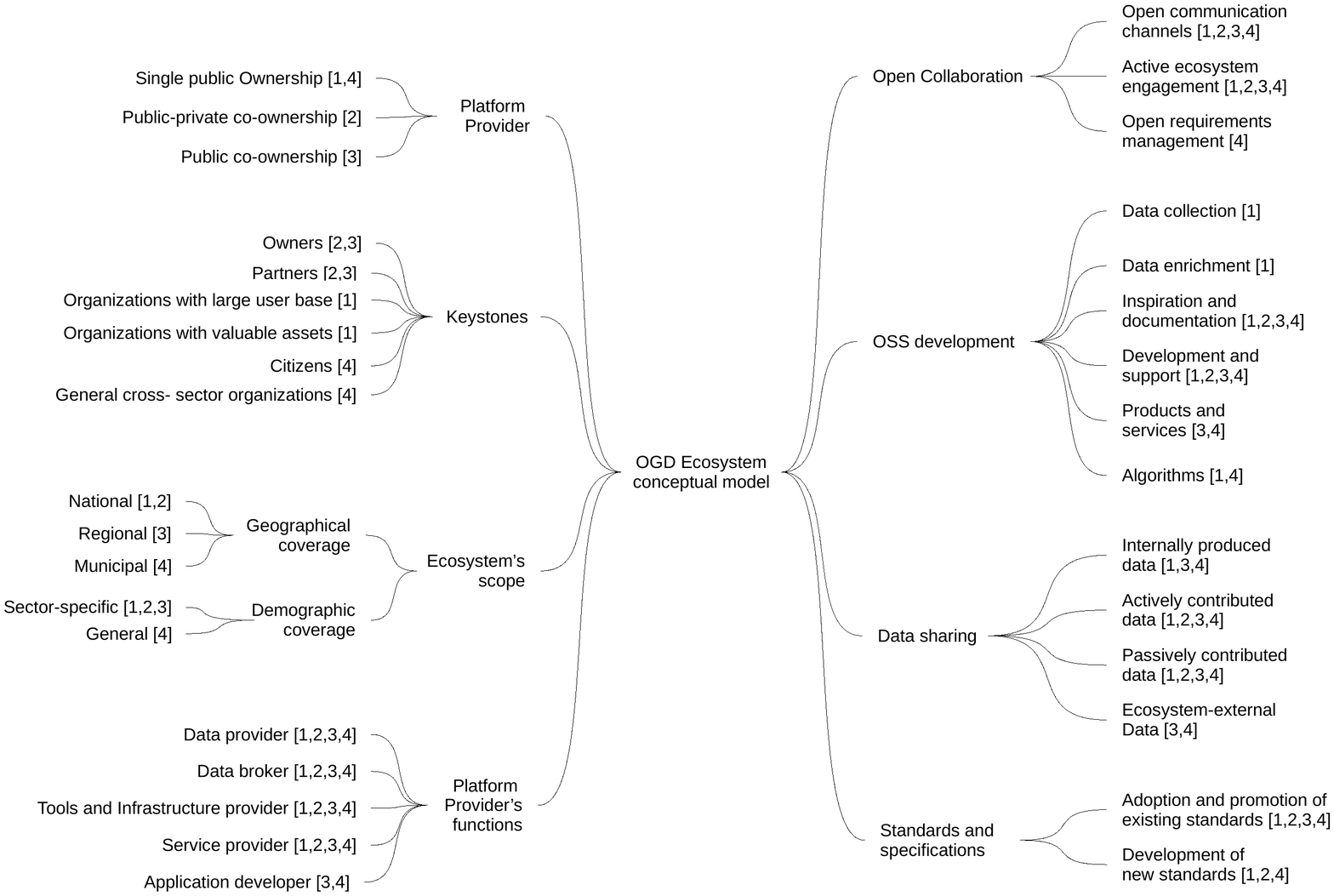}
\label{fig:collaborationFramework}
\end{sidewaysfigure}

JobTech Dev (Case 1), Trafiklab (Case 2), HSL DevCom (Case 3), and the City of Chicago (Case 4) present both similar and differentiating attributes as OGD ecosystems. In this section, we break these attributes down into several aspects, together forming a conceptual model emerging from the analysis (see Fig.~\ref{fig:collaborationFramework}) along with recommendations for platform providers to consider when (re)designing their OGD ecosystems to foster collaboration and sharing of resources. Each aspect of the model is explained and contrasted individually below, highlighting its different attributes that were identified in the four cases.


\subsection{Platform Provider and Keystones}

Among the studied cases, we observed three different types of platform providers based on the owner(s) it represents; \textit{Single-public ownership}, \textit{Public-private co-ownership}, and \textit{Public co-ownership}.
The \textit{single-public} ownership is represented by the case of JobTech Dev and the City of Chicago, which are governed by a single public entity, namely SPES and the city government, respectively. They decide who the keystones and main stakeholders are, and how they may influence the content and development of the platform (cf. keystone-based governance~\cite{dal2014role}). In SPES' case, they prioritized whom they consider large and important actors, while the City of Chicago maintains a broader definition of whom they consider as keystones. This freedom to choose allows them to potentially move quickly but also has the risk that their direction can easily change due to change in the national and local government, respectively. This risk could be a concern for both existing and new members in whether they can trust the direction and stability of the platform and ecosystem, and thereby if they should invest in platform integrations. 

The \textit{public-private} co-ownership model is represented by the case of Trafiklab, which is governed by Samtrafiken, an organization co-owned by all the Swedish public transport authorities along with a number of privately owned public transport actors.
The co-ownership of the platform provider provides a somewhat neutral body among the public and private actors, whom some may be considered as competitors. This consortium-based governance structure~\cite{dal2014role} may be a way to ensure the critical aspect of trust in the platform provider's commitment and a guarantee for the long-term stability and collaboration within the ecosystem~\cite{susha2020towards}. Similarities may be drawn to the role of data trusts~\cite{CoyleValue2020} as well as OSS foundations~\cite{de2013evolution} as proxy-organizations and neutral homes for data and OSS projects respectively. In these organizations, actors can collaborate and invest together in a way that benefits them all and with clear charters stating how the project will be technically governed. A risk with this form of ownership, however, is that the platform provider may not be able to receive government funding as it would benefit the private actors whom are part of its the ownership (in contrast to those who are not).

The \textit{public co-ownership} model is represented by the case of HSL DevCom, which is governed by the HSL, an organization co-owned by nine out of fourteen municipalities in the Helsinki Region. The remaining five have the option to join the collaboration should they prefer. This model allows for the municipalities to collaborate, as with Trafiklab, through a consortium-based governance structure~\cite{dal2014role} and also provide joint services and operations that are specific to their assignments as public entities.

\vspace{3mm}
\noindent\textbf{Recommendations: }Based on the studied cases, we advice public entities acting as platform providers in an OGD ecosystem to:
\begin{itemize}
    \item consider how to best enable collaboration and engagement from keystone players in the ecosystem through platform ownership. If the role of platform provider is concentrated to a single public entity (single-public ownership), it needs to enable dialogue and influence from keystones, for example, through an open collaboration. If ownership is distributed among the keystones, for example, a number of public entities (public co-ownership), or a number of public and private entities in collaboration (public-private co-ownership), collaboration may be ensured through formal organizational structures.
    \item identify and build trustful relationships with the keystones of the ecosystem as their engagement and success is important for the overall health and success of the ecosystem. What actors that constitute keystone will vary, but may include owners, partners, organizations with large user bases or other types of valuable assets, citizens, or general cross-sector organizations.
\end{itemize}


\subsection{Scope of Ecosystem}
The ecosystem can cover different scopes, both in terms of \textit{geographical} and \textit{demographic coverage}. The scope is in part defined by the platform provider who creates, governs, and facilitates the ecosystem. They set the vision and focus for the platform, and use cases it should address~\cite{iansiti2004keystone}. Returning to our definition of an OGD ecosystem (see Section~\ref{sec:introduction}), this connects to the common interest on which actors base their relationships. Continuing from this perspective, the scope is also partly defined by the actors who join, since without them there is no ecosystem. The vision and scope of the ecosystem is therefore something that potentially evolves organically with time.

Looking at geographical coverage, we can notice that JobTech Dev and Trafiklab have \textit{national} coverage of Sweden, while HSL DevCom and the City of Chicago have a more \textit{regional} and \textit{municipal} coverage on the municipal government. However, even though the main actors may be limited geographically, the same limitations do not have to concern all actors within the ecosystem. In the case of HSL DevCom, Google and Apple, for example, are interested in the public transport on a global level, not just in the Helsinki region in Finland.


Considering the demographic coverage, Trafiklab, HSL DevCom, and JobTech Dev have a \textit{sector-specific} focus on public transport and the Swedish labor market, respectively. However, even though the data may be limited, use cases don't have to be. These can be generic or positioned in a separate  domain where the data from the concerned platform is only one piece of the puzzle. In contrast, the City of Chicago has a very \textit{general} focus in terms of its ecosystem as the OGD and OSS they provide may relate to any part of the city's many services and functions.

\vspace{3mm}
\noindent\textbf{Recommendation: }Based on the studied cases, we advice public entities acting as platform providers in an OGD ecosystem to:
\begin{itemize}
    \item maintain a clear vision for the ecosystem, but at the same time allow for the ecosystem's actors to influence and buy in on the vision. It is important to also consider actors whose use case may stretch beyond the intended geographical and demographic coverage of the ecosystem.
\end{itemize}

\subsection{Platform Provider's Functions}
Looking at the functions, or responsibilities taken on by the platform providers in the investigated ecosystems, they cover many of the roles as reported in the literature~\cite{immonen2014requirements, lindman2015business}. Besides being \textit{data providers}, all four platform providers may be described as \textit{data brokers} as they gather, promote, and distribute data from third-party~\cite{immonen2014requirements} (cf. data intermediaries~\cite{susha2020towards}), but also data transformers as they transform data between different standards based on ecosystem needs~\cite{lindman2015business}. An example of the latter can be found in the case of Trafiklab where Samtrafiken transforms incoming data to the GTFS format.

Another important role that all platform providers take on is that of \textit{tool and infrastructure providers} as they develop supporting tools, frameworks, and example applications for their ecosystems, as well as infrastructure to enable collection and distribution of the data~\cite{immonen2014requirements}. 

Furthermore, all four platform providers also take on the role of \textit{service providers} as they to different extent provide services based on the data they publish. As examples, both HSL and Samtrafiken both provide their respective journey planners, the City of Chicago provides OpenGrid as a service to allow users to explore the OGD published on their data portal, and SPES enables individuals to store and distribute their resumé-data to recruitment platforms via their Career initiative. In the case of HSL and the City of Chicago, they also take on the role of \textit{application developers} as they develop and provide the applications (on which they base their services) as OSS.

Hence, to stimulate and enable collaboration in OGD ecosystems, the platform provider is required to perform multiple functions, which otherwise may be divided among multiple actors~\cite{immonen2014requirements, lindman2015business}. In contrast, when entering an existing ecosystem, an organization may consider taking a peripheral, less complex role~\cite{rudmark2019harnessing}. This is, however, a trade-off between influence on the platform development and value capture, i.e., if the organization's goals can still be achieved.

\vspace{3mm}
\noindent\textbf{Recommendation: }Based on the studied cases, we advice public entities acting as platform providers in an OGD ecosystem to:
\begin{itemize}
    \item consider the needs of the ecosystem's actors and take on an active and multi-functional role in order to stimulate and enable collaboration in the ecosystem. Such functions may include providing and brokering data between actors, as well as providing tools and infrastructure for adopting and processing the data, and related services and applications to enable use cases involving the data.
\end{itemize}

\subsection{Open Collaboration}
To enable an open collaboration within the ecosystem~\cite{linaaker2020public}, the platform provider itself needs to open up their processes and communication for the ecosystem using appropriate social and technical boundary resources~\cite{dal2014role, ghazawneh2013balancing}. 

\textit{Open communication channels} is one example of such resources through which the platform provider and ecosystem actors can collaborate and communicate with each other~\cite{feller2002understanding}. All four platform providers have such channels, either synchronous or asynchronous, complemented with alternative channels such as email, social media, and support desks. These channels usually serve as places for posting questions, asking for support but also reporting bugs, and requesting new functionality or data. 

\textit{Open requirements management} is another key boundary resource that includes openly communicating what requirements that the provider is working on~\cite{ernst2012case}. It also involves enabling ecosystem actors to comment as well as request features more formally, for example, through an issue tracker rather than the communication channels. None of the ecosystems has a central issue tracker but use those available via GitHub in connection to the OSS projects. In terms of communicating more long-term plans, only the City of Chicago publishes roadmaps related to their OSS projects.


\textit{Active ecosystem engagement} is the third type of social boundary resource, meaning that the platform providers need to communicate actively with their ecosystem actors, and also engage, initiate and facilitate collaborations within the ecosystem~\cite{linaaker2020public}. Beyond using the open communication channels, physical and virtual meetings such as hackathons, meetups, and conferences are commonly used among all the platform providers. This aspect is especially important before releasing a new data set or OSS project. For example, HSL reported struggles with gaining contributions as they have been too focused on the development and less on external engagement. As a counter-example one can consider the work that SPES has put into preparing the job-ad collaboration.



\vspace{3mm}
\noindent\textbf{Recommendation: }Based on the studied cases, we advice public entities acting as platform providers in an OGD ecosystem to:
\begin{itemize}
    \item enable an open collaboration within the ecosystem through the use of necessary and available social boundary resources, for example, by providing open communication channels, while also facilitating an open requirements management process and actively engaging the ecosystem in the collaboration. 
\end{itemize}

\subsection{OSS Development}
Among the studied cases, HSL has a stronger focus on OSS development as certain OSS projects have a more central role in the HSL DevCom ecosystem and are aimed for the end-users. In the other ecosystems, the OSS projects developed mainly serve as boundary resources to the developers in the ecosystems~\cite{dal2014role}. In the cases studied, we have identified six categories of OSS projects that platform providers may want to consider initiating and collaborating on: 

\begin{itemize}
    \item Data collection: In the case of JobTech Dev, SPES is employing OSS-based technology (developed and maintained by a third party supplier) to collect the job-ads from the different job-ad suppliers. 
    \item Data enrichment: OSS can also be used to enrich provided data with meta-data or in other ways add value to it. In the case of JobTech Dev, SPES provides internally developed functionality as OSS that helps actors to enrich job-ads with meta-data.
    \item Inspiration and documentation: All four platform providers develop and maintain application examples that can help users to understand how the data can be used. In this sense, the examples can be seen as a source of documentation. However, they may also inspire new use-cases that can be of value both for entrepreneurs and incumbents. 
    \item Development support: All four platform providers also see value in supplying common tools, libraries, and frameworks that can enable actors to create new products and services based on the data. 
    \item Algorithms: The City of Chicago publishes algorithms used within the city, 
    in part for transparency, but also to gain help in making the algorithms more effective and correct. SPES publishes its algorithms for external actors to test and validate the results they produce.
    \item Products and services: HSL takes it one step further and also develops OSS products that can be used both by service providers such as traffic operators and end-users, i.e., the commuters. 
\end{itemize}

\vspace{3mm}
\noindent\textbf{Recommendation: }Based on the studied cases, we advice public entities acting as platform providers in an OGD ecosystem to:
\begin{itemize}
    \item develop and collaborate with the ecosystem's actors on OSS that may improve adoption and innovation based on the data. For example, OSS that enables collection and enrichment of the data, provides inspiration and documentation, or general support to developers (such as tools, libraries, and infrastructure). The platform provider may also consider sharing their applications that use the data as input, which may include algorithms, as well as tools and consumer-facing products.
\end{itemize}

\subsection{Data Sharing}
To collaborate on data, the platform providers need to consider what type of data that is (or could be) shared and provided on the platform. If the data is produced externally, the platform provider takes on the role of a broker~\cite{immonen2014requirements} or intermediary~\cite{susha2020towards}, passing data on from the producers to the consumers which is mainly the case for Samtrafiken in the Trafiklab ecosystem. The other three ecosystems provide both internally and externally produced data. Differentiating between organizational structures, as proposed by Oliveira et al.~\cite{oliveira2019investigations}, therefore becomes difficult. Instead, to help characterize the types of data shared in an ecosystem, we have distilled four categories of data based on the studied cases; \textit{Internally produced} data, \textit{Actively contributed} data, \textit{Passively contributed} data, and \textit{Ecosystem-external} data.

Internally produced data originates from within the platform provider. In this case, the platform provider becomes the maintainer of the data and need to consider how they want actors to be able to contribute to it. In terms of the taxonomy data provided by SPES on the JobTech Dev platform, actors can mainly contribute through the established feedback channels. There are however examples where actors have contributed data sets directly to SPES which are then incorporated into the data set manually.

Actively contributed data means data that ecosystem actors actively contribute to the platform via tools or infrastructure provided by the platform provider. In this case, the contributing actors maintain their data and can themselves choose when to update what is available on the platform. Examples from the cases studied include the Career-initiative on the JobTech Dev platform, where individuals store their resumé-data, and city bike and car parking data, which is contributed to the HSL DevCom platform.

Passively contributed data means data that the platform provider collects from the ecosystem actors. The original data is maintained by the ecosystem actor, but what is available on the platform is dependent on the platform provider's collection and update procedures. One example is the traffic data from the regional transport authorities that are collected by Samtrafiken. A second example is the job-ads from the job-ad providers that are collected by SPES. A third example is the national data that is collected and broken down on a community level by the City of Chicago.

External ecosystem data refers to data that is maintained in another data ecosystem and collected by the platform provider and made available on the platform, in a direct or enriched way. One example concerns OpenStreetMap which map data is imported by HSL and used as input to their other data sets and API:s. If HSL or any of the ecosystem actors wants to contribute to the map data they can turn directly to the OpenStreetMap project.

\vspace{3mm}
\noindent\textbf{Recommendation: }Based on the studied cases, we advice public entities acting as platform providers in an OGD ecosystem to:
\begin{itemize}
    \item facilitate and enable the sharing of data, beyond their own. This can be done either by helping the actors to actively or passively contribute their data. In the active case, the actors use a technical infrastructure or communication channel provisioned by the platform provider. In the passive case, the platform provider collects the data from the actor to then publish it on the platform. A further aspect is to also consider how they can make use of external ecosystems to collaborate on data.
\end{itemize}

\subsection{Specification and Standardization}
The ecosystem enables a general \textit{adoption and promotion of existing standards} to promote interoperability among actors but also towards the surrounding or neighboring domains. In the studied cases, standards mentioned have mainly been domain-specific such as the GTFS format referring to traffic data (cf. vertical industry standards~\cite{rudmark2020open}). When the data producers use a different format, they transform it accordingly. Similarly, SPES has chosen to adopt the HROpen standard for their Career-initiative for individuals' resumés to be portable to different recruitment platforms internationally. When there are no existing standards that are suitable, the ecosystem offers a potential venue for collaborating on the \textit{development of new ones}, as in Trafiklab that developed a new standard for tickets and payment transactions.

\vspace{3mm}
\noindent\textbf{Recommendation: }Based on the studied cases, we advice public entities acting as platform providers in an OGD ecosystem to:
\begin{itemize}
    \item adopt and promote open standards where possible to enable interoperability within the ecosystem and well as to the outside. When needed, the platform provider should also take an active role in facilitating the development of new standards to achieve the same goal.
\end{itemize}



\section{Summary and Conclusions}
\label{sec:conclusions}
OGD ecosystems offers public entities new ways in how they can address their goals and directives. By opening up and enabling collaboration with cross-sectoral actors and the public through an ecosystem, the public entities can establish value networks with the potential to accelerate adoption and innovation based on the data. However, to exploit such benefits of open innovation, these public entities need to move beyond the role of data provider to that of a platform provider that actively enables and facilitates a collaboration on the data and related boundary resources within the ecosystem.


The research goal of this study has been to explore how public entities in the role as platform providers can enable collaboration in OGD ecosystems. To address this goal, we have conducted a multiple case study of four OGD ecosystems of varying characteristics in terms size, domain and geographical coverage. Based on the a cross-case analysis, we synthesize a conceptual model (see Fig.~\ref{fig:collaborationFramework}) that describe different aspects of OGD ecoystems and how these may vary. These aspects, for example, describes how collaboration can be enabled through different types of ownership of the platform provider, how the ecosystem's scope can vary, what roles the platform provider may undertake, how to enable open collaboration, and how to collaborate in terms of data sharing, OSS development, and standards. For each aspect, we provide recommendations that, along with the model, may provide guidance for public entities acting as platform providers can enable sharing and collaboration on both OGD as well as any boundary resources in the form of OSS and standards. 

As this study is limited to exploring the collaboration in four instances of OGD ecosystems, further research is required to validate both the model and recommendations, create a deeper understanding, and improve the external validity~\cite{runeson2012casestudy}. Readers should consider the context of the platform providers and their ecosystems as reported and adopt an analytical generalization to cases with similar characteristics~\cite{runeson2012casestudy}. 

In future research, we intend to continue our exploration into the collaborative aspects of OGD ecosystems in terms of both data, OSS, and standards, taking both the platform providers' and ecosystem actors' perspectives. We further aim to theorize how the data and software can be considered as commons and what role public entities can play in preserving sustainable provisioning and development of them.

\subsubsection{Acknowledgements} 
The research was funded through the JobTech Research Project, a collaboration between Lund University and SPES. The authors would like to thank the interviewees for their time and honesty, as well as the reviewers of the conference paper on which this study is based.

\bibliographystyle{splncs04}
\bibliography{references}
%

\end{document}